\documentstyle [12pt]{article}

\def\H{\mbox{I\hspace{-.15em}H\hspace{-.15em}I}}
\def\1{\mbox{I\hspace{-.15em}1}}

\def\Z{\mbox{Z\hspace{-.3em}Z}}

\font\goth=eufm10
\def\P{\hbox{\goth P}}

\def\R{{\rm I\hspace{-.15em}R}}

\def\C{\hspace{3pt}{\rm l\hspace{-.47em}C}}

\def\b{\begin{equation}} 
\def\e{\end{equation}}
\def\bee{\begin{enumerate}}
\def\eee{\end{enumerate}}

\title{Spin $\frac{1}{2}$ Field Theory in the\\ de Sitter space-time}
 
\author{ M. V. Takook \thanks{e-mail: takook@ccr.jussieu.fr}}
 
\date{\today}
\begin{document}
 
\maketitle {\it 
\centerline{ Laboratoire de Physique Th\'eorique et Math\`ematique}
\centerline{ Universit\'e Paris $7$ Denis-Diderot,75251 Paris Cedex 05, FRANCE}}

\begin{abstract}

A covariant quantization  
of the free spinor fields $(s=\frac{1}{2})$ in 
$4$-dimensional de Sitter (dS) space-time based on analyticity 
in the complexified pseudo-Riemanian manifold is presented. 
We define the Wigthman two-point function ${\cal W}(x,y)$, which satisfies the conditions
of: a) positivity, b) locality, c) covariance,  and d) normal analyticity. Then the Hilbert space structure and the field operators 
$\psi (f)$ are defined. A coordinate-independent formula 
for the unsmeared field operator $\psi (x)$ is also given. 
 
\end{abstract}

\vspace{0.5cm}
{\it Proposed PACS numbers}: 04.62.+v, 03.70+k, 11.10.Cd, 98.80.H
\vspace{0.5cm} 
 
\section{Introduction} 

In general on curved space-time, no true spectral condition can be 
satisfied by Quantum Field Theory (QFT) and no unique vacum state exists.
In the case of de Sitter spase-time, it has been discovered that the Hadamard condition selects an unique vacume state \cite{al}. The Hadamard condition is related to normal analyticity,
that is to say, the two-point function is the boundary value of an analytic function. One can replace the usual spectral condition by a certain geodesic spectral condition (or KMS condition), and one
can consider the generalized free fields on dS space-time. The generalized free fields can be defined entirely in terms of Wightman two-point function. In this context we present local free spinor fields (s=$\frac{1}{2}$) in 4-dimensional dS space-time based on analyticity in the complexified pseudo-Riemanian manifold. First we derive 
the dS-Dirac field equation as an eigenvalue equation for the 
Casimir operator and we find the solutions in terms of coordinate-independent dS plane-waves in tube domains.  
We define the two-point function $W(z_1,z_2)$ in terms of spinor
 dS plane-waves in their tube domains. Normal analyticity allows 
one to define Wigthman two-point function ${\cal W}(x,y)$ as 
the boundary value of $W(z_1,z_2)$ from the tube domains. Then 
the Hilbert space structure and the field operators 
$\psi (f)$ are defined. Finally, the unsmeared field operator $\psi (x)$ in terms of
a coordinate-independent dS plane waves is also defined. This work is in 
the continuation of the previous ones concerning the scalar 
case \cite{brgamo, brmo}.

\section{Notation}

De Sitter space-time is visualized as the hyperboloid with equation:
 $$ X_R=\{x^\alpha \in \R^5:x.x=\eta_{\alpha\beta}x^\alpha x^\beta=(x^0)^2-(x^1)^2-(x^2)^2-(x^3)^2-(x^4)^2 =-R^2\}$$
  \b \eta^{\alpha\beta}=\mbox{diag}(1,-1,-1,-1,-1);
       \;\;\alpha,\beta=0,1,...,4.\e
Let us define the punctured half-con with: $C=\{\xi^\alpha \in \R^5:\eta_{\alpha\beta}\xi^\alpha \xi^\beta=0\}$. The kinematical group of the de Sitter space-time is $G_R=SO_0(1,4)$ and the double (and universal) covering group of $G_R$ is given in a quaternionic realisation by \cite{taka}
\b Sp(2,2)=\{ g=\left( \begin{array}{clcr}  a &  b\\   c & d    \\ \end{array} \right)
  ,\;\;\hbox{det} g=1 ,\;\;\gamma^0\tilde g^t \gamma^0=g^{-1};\;a,b,c,d\in \H\},\e
$g^t$ denotes the $2 \times 2$ transpose of $g$, $\tilde g$  denotes its quaternionic conjugate,
 and  det$g$ is the determinant of $g$ viewed as a $4\times 4$ matrix with complex entries.
Now we need five $\gamma$ matrices instead of the usual four ones in Minkowski space-time. They are defined by the Clifford algebra:
 $$ \{ \gamma^\alpha ,\gamma^\beta \}=\gamma^\alpha \gamma^\beta+\gamma^\beta \gamma^\alpha =2 \eta^{\alpha \beta}\;\;\;,\;\;\;
  \gamma^{\alpha\dag}=\gamma^0 \gamma^\alpha \gamma^0.$$
The quaternion representation of the $\gamma$ matrices is given by \cite{tako}
$$ \gamma^0=\left( \begin{array}{clcr} \1 & \;\;0 \\ 0 &-\1 \\ \end{array} \right)
 ,\gamma^4=\left( \begin{array}{clcr} 0 & \1 \\ -\1 &0 \\ \end{array} \right)  $$
  \b   \gamma^1=\left( \begin{array}{clcr} 0 & i\sigma^1 \\ i\sigma^1 &0 \\    \end{array} \right)
     ,\gamma^2=\left( \begin{array}{clcr} 0 & -i\sigma^2 \\ -i\sigma^2 &0 \\  \end{array} \right)
      , \gamma^3=\left( \begin{array}{clcr} 0 & i\sigma^3 \\ i\sigma^3 &0 \\   \end{array} \right)\e
in terms of the $ 2\times2 $ unit $ \1 $ and Pauli matrices
$\sigma^i $. Casimir operator and infinitesimal generators read
\b  Q=-\frac{1}{2}L_{\alpha\beta}L^{\alpha\beta},\;\;
  L_{\alpha \beta}=M_{\alpha \beta}+S_{\alpha \beta}=-i(x_\alpha \bar\partial_\beta-x_\beta
         \bar\partial_\alpha)-\frac{i}{4}[\gamma_\alpha , \gamma_\beta],\e
where $\bar\partial$ is the tangential derivative:
$ \bar\partial_\beta = \partial_\beta + H^2 x_\beta x.\partial$.

\section{dS-Dirac field equation and plane-waves solution}

Starting from the Casimir operator and using the infinitesimal generators and the Casimir eigenvalue equation $ Q\psi(x)=(\nu^2+\frac{3}{2})\psi(x)$ \cite{mo} give
\b Q\psi(x)=\{(\frac{1}{2}\gamma_\alpha \gamma_\beta M_{\alpha \beta}+2i)^2+\frac{3}{2}\}
\psi(x)=(\nu^2+\frac{3}{2})\psi(x),\e
where $\psi(x)$ is the 4-component spinor wave function and $\nu \in \R$. A possible spinorial solution $\psi(x)$ to the above equation is afforded by the first-order equation \cite{di}:
 
 \b (-i \not x  \bar{\not \partial} +2i-\nu)\psi(x)=0,\;\;\; \mbox{(dS Dirac field equation)}\e
where $\not x=x.\gamma$ in usual notations.  For large $R$ behaviour $(R \longrightarrow \infty)$
we obtain the Dirac field equation in Minkowskian space. We know that any field quantity living on de Sitter space-time $X_R$ can be viewed as an homogenous function of the $\R^5$-variable $x^\alpha$ with some arbitrarily chosen degree $\sigma$ \cite{di}. Let us choose a solution to the dS-Dirac equation of the type \cite{tako}
\b \psi(x)=(\frac{1}{2}\gamma^\alpha \gamma^\beta M_{\alpha \beta}+i+\nu)\phi(x){\cal U}_T,\e
where $\phi(x)$ is a scalar field with degree $\sigma$ and ${\cal U}_T$ is an arbitrary four-component spinor. T denotes the ``orbital basis'' of $C^+$ with respect to a subgroup $L_e$ of $G_R$ which is the stabilizer of an unit vector $e$ ( $\mid e^2 \mid=1$) in $\R^5$ \cite{brmo}. We have two solutions for $\psi(x)$ \cite{tako}:
$$ \psi_1^{\xi,{\cal V}}(x)=(\frac{x.\xi}{R} )^{-2+ i \nu}
\frac{\not x \not {\xi}}{ R\sqrt{2(\xi^0+1)}} {\cal
U}_T(\stackrel{o}{\xi}_+) \equiv(\frac{x.\xi}{R} )^{-2+ i \nu}{\cal
V}(x,\xi),$$
 \b \psi_2^{\xi,{\cal U}}(x)=(\frac{x.\xi}{R} )^{-2- i \nu} \frac{ \not {\xi}\gamma^4}
 {\sqrt{2(\xi^0+1)}}{\cal U}_T(\stackrel{o}{\xi}_-)\equiv(\frac{x.\xi}{R} )^{-2- i \nu}{\cal U}(\xi).\e
We now consider the transformation of the Dirac free field $\psi(x)$ in such a way that the
transformd $\psi'(x')$ obyes the same dS-Dirac equation in the new frame . This lead to the simple relation  $\psi'(x)=g\psi(\Lambda^{-1}(g)x)$ where $ g \in Sp(2,2)$  is viewed as a $4\times 4$ matrix when acting on the 4-spinor $\psi(x)$ and $\Lambda \in SO(1,4)$. We have 
\b \psi'^{\xi,{\cal U}_T}(x)=g \psi^{\xi,{\cal U}_T}(\Lambda^{-1}x)=\psi^{\Lambda\xi,g{\cal U}_T}(x).\e
We see how the dS action is transfered onto the "reciprocal space " $C^+$ to which the parameter $\xi$ belongs, and the 4-component arbitary spinor ${\cal U}_T$. The plane-wave solutions to the free Dirac equation in Minkovskian space are given by the large-R behavior of the plane-wave solutions to the dS-Dirac field equation \cite{tako}. For obtaining general field solutions, we consider the solution in the compolexed dS space-time $X_R^{(c)}$
$$ X_R^{(c)}=\{ z=x+iy\in  \C^5;\;\;\eta_{\alpha \beta}z^\alpha
z^\beta=(z^0)^2-\vec z.\vec z-(z^4)^2=-R^{2}\}$$ 
\b =\{ (x,y)\in
\R^5\times  \R^5;\;\; x^2-y^2=-R^{2},\; x.y=0\}.\e 
Let $T^\pm= \R^5+iV^\pm$ be the forward and backward tubes in $ \C^5.\;
V^+$(resp. $V^-)$ stems from the causal structure on $X_R$, 
\b V^\pm=\{ x\in \R^5;\;\; x^0\stackrel{>}{<} \sqrt {\parallel \vec
x\parallel^2+(x^4)^2} \}.\e 
We then introduce their respective
intersections with $X_R^{(c)}, \;\;\;{\cal T}^\pm=T^\pm\cap X_R^{(c)}$,
which will be called forward and backward tubes in the complex dS space-time
$X_R^{(c)}$. Finally we define the set 
$$ {\cal T}_{12}=\{ (z_1,z_2);\;\;
z_1\in {\cal T}^+,z_2 \in {\cal T}^- \}, $$
as a tube above $X_R\times X_R$ in $X_R^{(c)}\times X_R^{(c)}$. Details are given in \cite{brmo}. When $z$ varies in ${\cal T}^+$ (or
${\cal T}^-$) and $\xi$ lies in the positive cone ${\cal C}^+$,
the plane wave solutions $\psi^{\xi,{\cal
V}}(z)=(\frac{z.\xi}{R} )^\sigma {\cal U}(z,\xi),\; \sigma \in \C$ are globally
defined because the imaginary part of $(z.\xi)$ has  a fixed sign. Now we can define the Wightman two-point function.

\section{Two point function}

Let us briefly recall the conditions we require on the Wightman two-point function 
$$ {\cal W}^{\nu}(x,y)=<\Omega,\psi(x)\bar \psi(y) \Omega >,$$  
where $x,y \in X_R$ and 
$  \bar\psi=\psi^\dagger\gamma^0\gamma^4,$
 is the spinor field conjugate to $\psi$ . This function is $4\times 4$ matrix-valued in the present case, and has to satisfy the following requirements:
\begin{enumerate}
\item[a)] {\bf Positivity}

for any test function
$f \in {\cal D}(X_R)$ with values in $\C^4$
\begin{equation} \int _{X_R \times X_R}  \bar f(x){\cal W}(x,y)
f(y)d\sigma(x)d\sigma(y)\geq       0,\end{equation}
where  $d\sigma (x)$ denotes the dS-invariant measure on $X_R$. 
 
\item[b)] {\bf Locality} 

for every space-like separated pair $(x,y)$, {\it i.e.} $x\cdot y>-R^2$,
\b {\cal W}_{i\bar j}(x,y)=-{\cal W}_{\bar j i}(y,x),\e
where ${\cal W}_{\bar ji}(y,x)=<\Omega,\bar \psi_{\bar j}(y)
\psi_i(x) \Omega >$.
 \item[c)] {\bf Covariance}
     \b g{\cal W}
(\Lambda^{-1}(g)x,\Lambda^{-1}(g)y)i(g^{-1})={\cal W}(x,y) , \e
where $i(g^{-1})=-\gamma^4 g^{-1} \gamma^4$ \cite{tako},

\item[d)] {\bf Normal analyticity}

${\cal W}(x,y)$ is the boundary value (in the sense of distributions) of a function
$W(z_1,z_2)$ which is analytic in the domain ${\cal T}_{12}$.

\end{enumerate}

The two-point $W^\nu(z_1,z_2)$ , lablled by the principal-series parameter $\nu$, is given by the following class of integral representations \cite{tako}
$$ W^\nu_{i\bar j  } (z_1,z_2)=c_\nu \int_T
(z_1.\xi)^{-2-i\nu}(z_2.\xi)^{-2+i\nu}\sum_{a=1,2}{\cal
U}^a_i(\xi)\bar{\cal U}^a_{\bar j}(\xi)d\mu_T(\xi),$$
and the two-point function $W^\nu_{\bar j i } (z_2,z_1)$ is given by the
following class of integral representations 
$$ W^\nu_{\bar j i }
(z_2,z_1)=H^2c_\nu \int_T (z_1.\xi)^{-2+i\nu}(z_2.\xi)^{-2-i\nu}
  \sum_{a=1,2}{\cal V}^a_i(z_1,\xi)\bar{\cal V}^a_{\bar j}(z_2,\xi) d\mu_T(\xi).$$
$d\mu_T(\xi)$ is a measure invariant under $ L_e$. We can write
 \b
W^\nu(z_1,z_2)=D(z_2)\gamma^4 {\cal N}(z_1,z_2),\;\;D(z_2)=\frac{1}{\nu+i}(-i\not z_2\not{\bar{\partial}}_{z_2}+i+\nu).$$
${\cal N}(z_1,z_2)$ is a scalar two-point function
$$  {\cal N}(z_1,z_2)=c_{\nu } \int_T(z_1.\xi)^{-2-i\nu}   (z_2.\xi)^{-1+i\nu}d\mu_T(\xi).\e
In terms of the generalized Legendre function of the first kind we
have \cite{brmo, tako} 
$$ W^\nu_{i\bar j}
(z_1,z_2)=C_{\nu}(D(z_2)\gamma^4)_{i\bar j}
P_{-1+i\nu}^{(5)}(\frac{z_1.z_2}{R^2})=4C_{\nu}(D(z_2)\gamma^4)_{i\bar j}(1-\frac{z_1.z_2}{R^2})^{-\frac{1}{2}}\P^{-1}_{i\nu}(\frac{z_1.z_2}{R^2})$$
\b W^\nu_{\bar j i }
(z_2,z_1)= 4H^2C_{\nu}(\not z_1D(z_1)\not z_2 \gamma^4)_{i\bar
j}(1-\frac{z_1.z_2}{R^2})^{-\frac{1}{2}}\P_{i\nu}^{-1}(\frac{z_1.z_2}{R^2}).\e 
The boundary value of $W^\nu_{i\bar j}
(z_1,z_2)$ provides us with the following represntation for the Wightman two-point function \cite{tako}: 
$$ {\cal W}(x,y)=c_{\nu}\int_T [(x.\xi)_+^{-2-i\nu}+e^{i\pi(-2-i\nu)} (x.\xi)_-^{-2-i\nu}]$$
\b[(y.\xi)_+^{-2+i\nu}+e^{-i\pi(-2+i\nu)}(y.\xi)_-^{-2+i\nu}] \not
{\xi} \gamma^4 d\mu_T.\e 
This function satisfies the conditions
of: a) positivity, b) locality, c) covariance,  and d) normal analyticity \cite{tako}. The existence of ${\cal W}$ allows one to make the QF formalism work \cite{stwi}. 
The spinor fields $\psi(x)$ are expected to be operator-valued distributions on $X_R$ acting on a Hilbert space ${\cal H}$ . The Hilbert space ${\cal H}$  of the representation can be described as the Hilbertian sum 
$$ {\cal H}={\cal H}_0\bigoplus[\bigoplus_{n=1}^{\infty}A{\cal
H}_1^{\bigotimes n}].$$
$A$ denotes the antisymmetrisation
operation and ${\cal H}_0=\{ \lambda \Omega,\;\; \lambda \in
\C\}$ where the vector $\Omega$, cyclic for the polynomial algebra of field operators and invariant under the representation of $G_R$, is ``the vacuum" . ${\cal H}_1$ is defined by the scalar product \b
(h_1,h_2)=\int_{X_R \times X_R} \bar h_1(x){\cal
W}(x,y)h_2(y)d\sigma(x)d\sigma(y)\geq 0,\e where $h \in {\cal
D}(X_H)$ with values in $\C^4$. Each field operator $\psi(f)$ can be defined in terms of annihilation and creation operators by \cite{tako}:
$$\psi(f)h^{(n)}(i_1,
x_1;i_2, x_2;...;i_n ,x_n)=$$ $$\sqrt{n+1} \int_{ X_R \times X_R }
f^i(x)  {\cal W}_{\bar j i}(y,x)h^{(n+1)}(\bar j,y;i_1,
x_1;...;i_n ,x_n)d\sigma(x)d\sigma (y)$$
     \b +\frac{1}{\sqrt{n}}\sum_{k=1}^n(-1)^{k+1} f_{i_k}(x_k)h^{(n-1)}
     (i_1,x_1;...;\hat{i_k}, \hat{x_k};...;i_n, x_n),\e
where the $i_k=1,2,3,4,$ are spin indices. Here `$\hat{i_k}$'
means omit it. By using the Fourier-Bros transformation on $X_R$ \cite{brmo}, we can write the unsmeared field operators $\psi(x)$ \cite{tako} 
$$ \psi(x)=\int_T \sum_{a=1,2} \{\; a_a(\xi,\nu){\cal
U}^a(\xi)[(x.\xi)_+^{-2-i\nu}+e^{i\pi(-2-i\nu)}(x.\xi)_-^{-2-i\nu}]
$$ \b  +d^{\dag}_a(\xi,\nu){\cal V}^a(x,\xi)[(x.\xi)_+^{-2+i\nu}+e^{-i\pi(-2+i\nu)}(x.\xi)_-^{-2+i\nu}]\;
\} d\mu_T(\xi),\e where $a_a(\xi,\nu)$ and $d_a(\xi,\nu)$ are
defined by $ a_a(\xi,\nu)\mid \Omega>=0=d_a(\xi,\nu)\mid \Omega>$. 
The field anticommutator is given by
 $$\{\psi_i(x),\bar \psi_{\bar j}(y)\}={\cal W}_{i\bar j}(x,y)+{\cal W}_{\bar j i}(y,x).$$
It can be esily checked that for space-like separated points (x,y) we have $\{\psi_i(x),\bar \psi_{\bar j}(y)\}=0$ \cite{tako}. The integral representation for the two-point function is defined on the space which carries the principal series of the dS group $Sp(2,2)/ \Z_2 \approx SO_0(1,4)$. In the limit $R \longrightarrow \infty$, we obtained \cite{tako}
 $$ \lim_{R \longrightarrow \infty}\{\psi(x),\bar \psi(y)\}=$$
\b \frac{1}{2(2\pi)^3} \int \{e^{-ik.(X-Y)} (\not k \gamma^4+m)+
e^{ik.(X-Y)}(\not k \gamma^4-m)\}\frac{d^3k}{k^0},\e which is of
the same form of the Minkowskian space.

\section{Conclusion and outlook}

Using the dS-plane waves in tube domains for spinor field, one
can construct an analytic function $W(z_1,z_2)$ that its boundary
value is the Wightman two-point function ${\cal W}(x,y)$.
Then the dS-plane-waves allow us to construct the quantum
field on dS space in the same way as the quantum field on
Minkowski space. In the case of the massless spinor field, we must
replace $\nu$ with $0$ in the  Waightman two-point function as well as in the field operator $\psi(x)$ \cite{tako}. In this case the corresponding UIR is known as the first term of
the spinor discrete series of representation, which is written as
$\Pi_{\frac{1}{2},\frac{1}{2}}^\pm$ in \cite{dix}. The
$\pm$ define the helicity of the massless spinor field. 
We now intend to use these methods to
construct a covariant quantum fields with spin-1 and spin-2. In
the case of the massive field the procedure is the same as
spin-$\frac{1}{2}$. In the case of the massless field, we must
use the Gupta-Bleuler quantization for obtaining a fully covariant
theory. At this moment, this kind of work is in progress on the "massless" representations of dS group and the related QFT. 

\vskip 0.5 cm
 
 \noindent {\bf{Acknowlegements}}: I am grateful to Professor J. P.
 Gazeau for giving inspiration to this investigation and for a fruitful discussion about it.

\end{document}